\documentclass[%
 aip, apl,
 amsmath,amssymb,
 reprint,%
]{revtex4-1}

\usepackage{graphicx}
\usepackage{dcolumn}
\usepackage{bm}
\usepackage{xcolor}

\usepackage{soul}

\usepackage[utf8]{inputenc}
\usepackage[T1]{fontenc}
\usepackage{mathptmx}
\usepackage{etoolbox}

\makeatletter
\def\@email#1#2{%
 \endgroup
 \patchcmd{\titleblock@produce}
  {\frontmatter@RRAPformat}
  {\frontmatter@RRAPformat{\produce@RRAP{*#1\href{mailto:#2}{#2}}}\frontmatter@RRAPformat}
  {}{}
}%
\makeatother
\begin{document}

\preprint{AIP/123-QED}

\title{Regular sloshing modes in irregular cavities using metabathymetry}
\author{Adam Anglart}
    \affiliation{PMMH, ESPCI, Universit\'e PSL, Sorbonne Universit\'e, 1 rue Jussieu, 75005 Paris, France}
    \email{adam.anglart@espci.fr}
\author{Agn\`es Maurel} 
    \affiliation{Institut Langevin, ESPCI, 1 rue Jussieu, 75005 Paris, France}
\author{Philippe Petitjeans}  
    \affiliation{PMMH, ESPCI, Universit\'e PSL, Sorbonne Universit\'e, 1 rue Jussieu, 75005 Paris, France}
\author{Vincent Pagneux}  
    \affiliation{LAUM, Universit\'e du Maine, Avenue Olivier Messiaen, 72085 Le Mans, France}

\date{\today}

\begin{abstract}
We present a comprehensive investigation, combining numerical simulations and experimental measurements, into the manipulation of water waves and resonance characteristics within closed cavities utilizing anisotropic metamaterials. We engineer the anisotropic media with subwavelength-scale layered bathymetry through the application of coordinate transformation theory and the homogenization technique to a fully three-dimensional linear water wave problem. Experimental and numerical analyses of deformed cavities employing anisotropic metamaterial bathymetry demonstrate regular sloshing mode patterns and eigenfrequencies akin to those observed in rectangular reference cavities with flat bathymetry. Our study underscores the potential of water wave metamaterials for establishing robust anisotropic metabathymetry for the precise control of sloshing modes.
\end{abstract}

\maketitle

\begin{figure}[b] 
  \includegraphics[width=.35\textwidth]{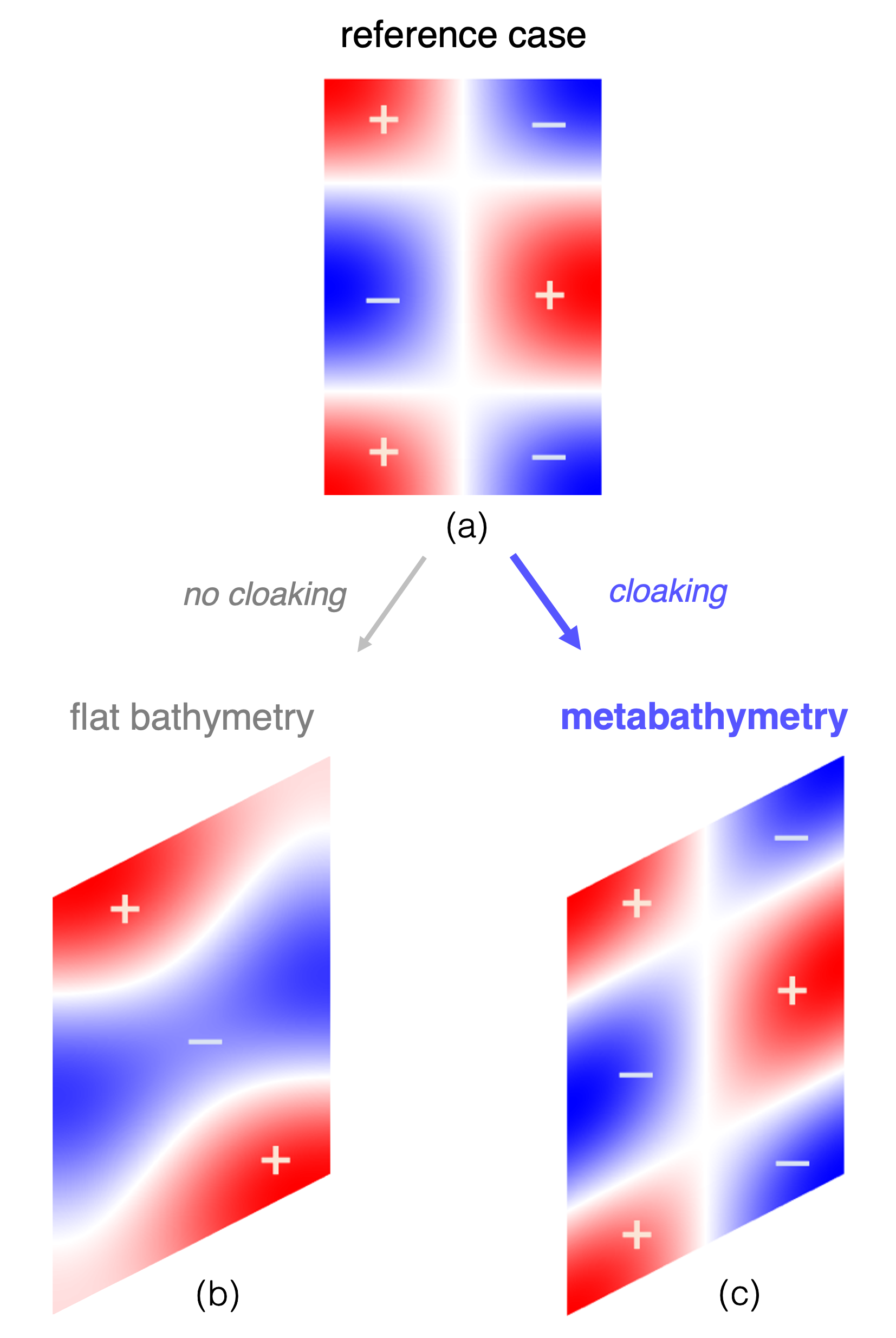}
   \caption{\label{fig:principe} Example of a regular (rectangular) cavity eigenmode (a) and its equivalent in an irregular cavity without (b) and with cloaking (c).}
\end{figure}  


The pursuit of designing materials with extraordinary properties, diverging from those commonly found in nature, has captured significant attention in the scientific and engineering community. Metamaterials, originally used in electromagnetism \cite{ziolkowski2006metamaterials,cui2010metamaterials}, have been of substantial interest in controlling acoustic \cite{achilleos2017non,faure2016experiments,fan2020reconfigurable,ma2016acoustic,qi2016acoustic}, elastic \cite{nassar2017non,dong2017topology,trainiti2019time,xu2020physical}, and seismic waves \cite{colombi2016seismic, miniaci2016large, achaoui2017clamped,maurel2018conversion,pham2020hybridized}. They have also been used to control water wave propagation, for which some applications include cloaking \cite{newman2014cloaking,bobinski2018backscattering, farhat2008broadband, alam2012broadband, zhang2020invisibility, dupont2015numerical, zhang2023wavelength}, focusing \cite{bobinski2015experimental, li2018concentrators}, or guiding the energy flow \cite{berraquero2013experimental, wang2017carpet}. As a result, metamaterials can be beneficial for coastal engineering, when it comes to wave-free zones, shore protection, energy harvesting, or designing wake-less watercraft. However, sloshing dynamics, which is crucial for various applications in fluid containment and transport, fluid-structure interaction, tuned mass damping, and structural integrity in marine and engineering contexts \cite{ibrahim2001recent, frandsen2004sloshing, ibrahim2005liquid}, has received comparatively less attention from the point of view of metamaterial design. Such dynamics, influenced by cavity geometry, gives rise to a range of phenomena such as sloshing resonance and high spots \cite{kulczycki2011high,kulczycki2016shape}. In this paper, by leveraging the coordinate transformation theory \cite{pendry2006controlling,leonhardt2009transformation} and the homogenization approach to the fully three-dimensional water wave problem \cite{rosales1983gravity,Maurel2017}, we explore theoretically and experimentally the possibility of controlling sloshing modes through carefully designed immersed metabathymetry. The idea is to use the coordinate transformation theory (CTT), classically applied in the scattering context for cloaking, in a closed cavity in order to mimic the resonance of a regular separable geometry both for the patterns and the eigenfrequencies (Fig.~\ref{fig:principe}) \cite{anglart2020regular, Anglart2021}. In other words, we aim at "cloaking" eigenmodes.

\begin{figure}[t]
                \includegraphics[width=.48\textwidth]{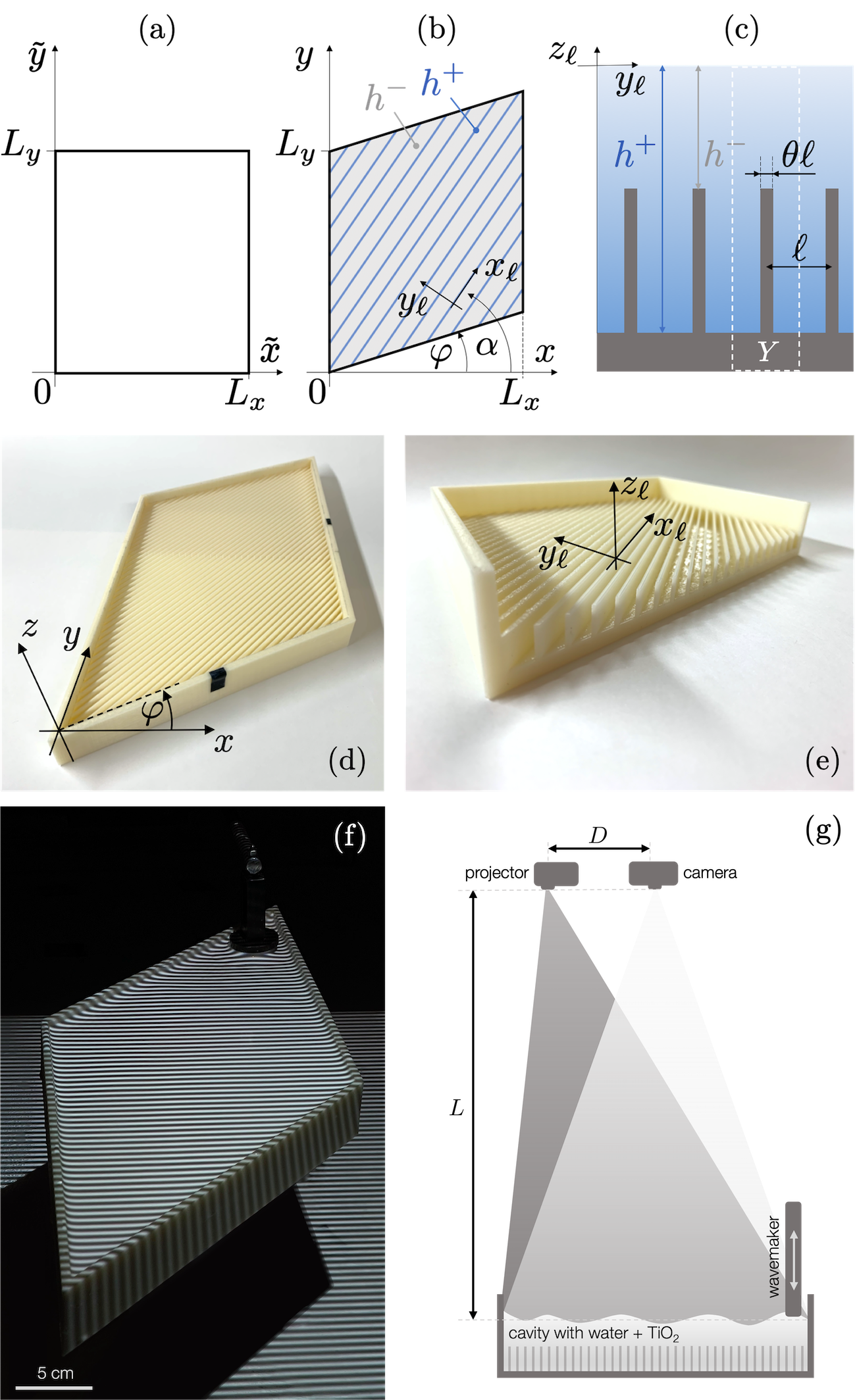} 
                \caption{\label{fig:photoExp} The cavity in virtual space $(\tilde x,\tilde y)$ (a) and in the real space $(x,y)$ with shear angle $\varphi$ (b). In (b) we show the top view of the metabathymetry realizing the anisotropy, an array of layers inclined by an angle $\alpha$ and in (c) the side view. (d) Metamaterial cavity - the view from above with a coordinate system in real space $(x,y,z)$. (e) Cut view with a local coordinate system used for metabathymetry design $(x_\ell, y_\ell, z_\ell)$. (f) Picture of the experimental setup with the cavity filled with water with dissolved titanium dioxide. The fringe pattern is projected onto the water surface. The wavemaker (linear motor) is placed in the top right corner of the cavity. (g) Scheme of the experimental setup.}
                \end{figure}

We start with the homogeneous two-dimensional shallow-water equation (2D SWE) \cite{mei1989applied, mei2005theory} for a closed cavity with vertical walls in a virtual space $(\tilde x, \tilde y )$ (Fig.~\ref{fig:photoExp}a) 
                \begin{align}
                \tilde \nabla \cdot \left( h_0 \tilde \nabla \eta \right) + \frac{\omega^2}{g} \eta = 0, \nonumber \\ 
                \mathbf{n}   \cdot  \tilde \nabla \eta = 0  \text{ on walls},
                \label{SWEvirtualSpace}
                \end{align} 
where $h_0$ is the constant water depth, $\eta$ stands for the vertical displacement of the free surface, $\omega$ is the frequency, $g$ denotes the gravitational acceleration, $\mathbf{n}$ identifies the vector normal to the boundary, and $\tilde  \nabla=(\partial/\partial \tilde x, \, \partial/\partial \tilde y)^T$. The shallow-water limit assumes that the wavenumber $kh_0\ll 1$, where $k$ is given by the dispersion relation $k=\omega/\sqrt{gh_0}$. The problem in Eq.~\eqref{SWEvirtualSpace} describes the eigenvalue problem (eigenvalue $\omega$) of sloshing modes in a closed cavity. In a regular, rectangular cavity with dimensions $L_x$ and $L_y$ (Fig. 2a), the solution to the eigenvalue problem (1) yields a discrete set of eigenfunctions $ \eta_{nm}(\tilde x, \tilde y) = \cos\left(n\pi \tilde x/L_x\right)\cos\left(m\pi \tilde y/L_y\right)$ (Fig. 1a), where $n, m = 0, 1, 2, \ldots$, and corresponding eigenvalues (eigenfrequencies) are given by $\smash{k_{nm} = \sqrt{(n\pi/L_x)^2 + (m\pi/L_y)^2}}$. In a deformed cavity with flat bathymetry (Fig. 1b), the eigenmodes and eigenfrequencies differ significantly from those in a regular, rectangular cavity and they usually must be computed numerically.
        
Our objective is to achieve the same eigenfrequencies and eigenmodes of a regular cavity ($k_{nm}$, $\eta_{nm}$)  in a cavity with irregular geometry as shown in Fig \ref{fig:principe}c. To do so, we employ coordinate transformation theory with the following mapping $x = \tilde x$, $y = \tilde x \tan \varphi + \tilde y$ \cite{berraquero2013experimental, chen2008electromagnetic}. After applying such deformation (Fig. \ref{fig:photoExp}), the equations \eqref{SWEvirtualSpace} in $(\tilde x, \tilde y)$ coordinates, governing the wavefield in the regular cavity, become
                \begin{align} 
                \nabla \cdot \left( \mathbf{J} h_0 \mathbf{J}^T \nabla \eta \right) + \frac{\omega^2}{g } \eta = 0, \nonumber \\
                \mathbf{n} \cdot \left( \mathbf{J} h_0 \mathbf{J}^T \nabla \eta \right) = 0 \quad \text{on walls}, 
                \label{SWErealSpace}
                \end{align} 
in $( x,  y)$ coordinates, governing the wavefield in the irregular cavity, where $\nabla=(\partial/\partial  x, \, \partial/\partial  y)^T$ and $\mathbf{J}$ represents the Jacobian matrix of the transformation, namely
                \begin{equation}
                \mathbf{J}= \begin{bmatrix} 
                        1 & 0 \\
                        \tan \varphi & 1  
                \end{bmatrix}.
                \label{Jacobian}
                \end{equation}

By carefully designing a bathymetry that enforces an anistropic effective medium equations identical to \eqref{SWErealSpace} with effective water depth tensor $\mathbf{h}_\text{eff}=\mathbf{J} h_0 \mathbf{J}^T$, both the irregular and regular cavities will share the same eigenfrequencies $k_{nm}$ and eigenmodes, i.e., $\eta_{nm} (x,  y)~=~ \cos\left(n\pi  x/L_x\right)\cos\left(m\pi ( y- x \tan \varphi )/L_y\right)$ (Fig. 1c). To practically realize $\mathbf{h}_\text{eff}$ in Eq. \eqref{SWErealSpace}, we use results given by 3D homogenization of layered bathymetry \cite{rosales1983gravity, bensoussan2011asymptotic,Maurel2017}, which provides
                \begin{align}
                \mathbf{h}_\text{eff}=\mathbf{R}_\alpha \mathbf{h} \mathbf{R}_\alpha^T \quad \text{with} \quad \mathbf{h} = \begin{bmatrix} 
                    h_{x_\ell} & 0 \\  
                    0 & h_{y_\ell} 
                \end{bmatrix},
                \label{SWEanisotropy}
                \end{align} 
and with $\mathbf{R}_\alpha$ representing the rotation matrix through an angle $\alpha$ with respect to the $x$ axis (Fig. \ref{fig:photoExp}b). This anisotropic medium \eqref{SWEanisotropy} is characterized by distinct effective water depths in the $x_\ell$ and $y_\ell$ directions (Fig. \ref{fig:photoExp}b), denoted as $h_{x_\ell}$ and $h_{y_\ell}$ respectively. Eventually, identifying the anistropic effective medium tensor \eqref{SWEanisotropy} with the geometrical transformation tensor \eqref{SWErealSpace} allows to write that $ \mathbf{J} h_0 \mathbf{J}^T=\mathbf{R}_\alpha \mathbf{h}  \mathbf{R}_\alpha^T$, which yields explicit formulas for water depths $(h_{x_\ell}, h_{y_\ell})$ and rotation angle $\alpha$ in terms of $\varphi$ and $h_0$ \cite{Anglart2021}. Specifically, the depths $h_{x_\ell}$ and $h_{y_\ell}$ satisfy the equation $h^2-(2+\tan^2\varphi)h_0 h+h_0^2=0$, where $h_{x_\ell}>h_{y_\ell}$ and $\tan 2\alpha=-2/\tan\varphi$ (note a misprint in \cite{berraquero2013experimental}). The boundary information is encoded within the metabathymetry as a result of the homogenization process. The condition is expressed as $ \mathbf{n} \cdot \left( \mathbf{R}_\alpha \mathbf{h} \mathbf{R}_\alpha^T \nabla \eta \right) = 0 $ on walls \cite{homogenization2024}.

In the preceding analysis, we obtained the values of $h_{x_\ell}$ and $h_{y_\ell}$ for a given shear angle $\varphi$ and a reference water depth $h_0$ purely by means of the geometrical transformation. Experimental realization of the effective anisotropic medium can be achieved by incorporating stratified, layered thin vertical plates as a bathymetry of the cavity\cite{Maurel2017}, as shown in Fig.~\ref{fig:photoExp}c. In this paper, we call this layered structure a \textit{metabathymetry}. Our focus now shifts towards determining the dimensions of the metabathymetry $h^+$, $h^-$, $\theta$, and $\ell$ (Fig.~\ref{fig:photoExp}c) that would fulfill the anisotropy given by the already established values of $h_{x_\ell}$ and $h_{y_\ell}$ as in Eq.~\eqref{SWEanisotropy}. It has been shown that effective anisotropy ($h_{x_\ell}$, $h_{y_\ell}$), being the result of the presence of the metabathymetry ($h^+$, $h^-$), cannot be inferred from the shallow-water equation even in the shallow water regime, as 3D effects affect the flow over a rapidly varying bathymetry due to the strong effect of the evanescent field \cite{rosales1983gravity,Maurel2017}. This is why, to properly model the aforementioned effects, we use the homogenization of the full 3D linear water wave problem. Considering a harmonic regime with time dependence $\mathrm{e}^{-\mathrm{i}\omega t}$, assuming that the fluid is inviscid and incompressible, and knowing that the flow is irrotational, the velocity potential $\phi(x_\ell,y_\ell,z_\ell)$ satisfies
                \begin{align}
                \bigtriangleup \phi = 0, \nonumber \\ 
                \frac{\partial \phi}{\partial z_l} = \frac{\omega^2}{g}\phi \text{ at } z_l=0, \nonumber \\
                \mathbf{n} \cdot \nabla\phi=0 \text{ on } \Gamma,
                \label{3DLaplace}
                \end{align}
where $\Gamma$ is the nonflat bathymetry, $\mathbf{n}$ is the vector normal to it, $z_l=0$ corresponds to the unperturbed free surface.

As shown in \cite{Maurel2017,rosales1983gravity}, the effective water depths $h_{x_\ell}$ and $h_{y_\ell}$ derived from a fully three dimensional problem \eqref{3DLaplace} are of the form
                \begin{align}
                h_{x_\ell} = \ell \int_Y \frac{\partial \Phi}{\partial x_l} \mathrm{d}Y, \quad h_{y_\ell}=\langle h \rangle,
                \label{3D}
                \end{align}
where $\Phi$ represents the potential satisfied in the unit cell $Y$, $\ell$ is the periodicity of the metabathymetry, and $\langle h \rangle = \theta h^- + (1-\theta)h^+$, where $\theta$ is the filling fraction of the layers \cite{Maurel2017} (Fig.~\ref{fig:photoExp}c). Now, we use the homogenization of a fully three-dimensional linear water wave problem \eqref{3D} to calculate the water depths $\mathbf{h}^\pm =[h^+ \ h^-]^T $. However, this approach needs already preset dimensions of the metabathymetry creating an inverse problem. In order to find $\mathbf{h}^\pm$ we solve the problem of optimization with constraints using the Nelder-Mead simplex search method\cite{lagarias1998convergence} available in Matlab Optimization Toolbox: $\min_{\mathbf{h}^\pm \in \mathbb{R}^2} \left| \left| f(\ell, \theta, h^+, h^-) - [h_{x_\ell} \ h_{y_\ell}]^T\right| \right|$, subject to $h^+-h^- \geq 0$, where $f(\ell, \theta, h^+, h^-)$ is a solution of \eqref{3D}, i.e., the values of $h_{x_\ell}$ and $h_{y_\ell}$, obtained with the use of multimodal method\cite{Maurel2017}. The periodicity $\ell$ and the filling fraction $\theta$ are constant.
                \begin{figure}[t]
                \includegraphics[width=.4\textwidth]{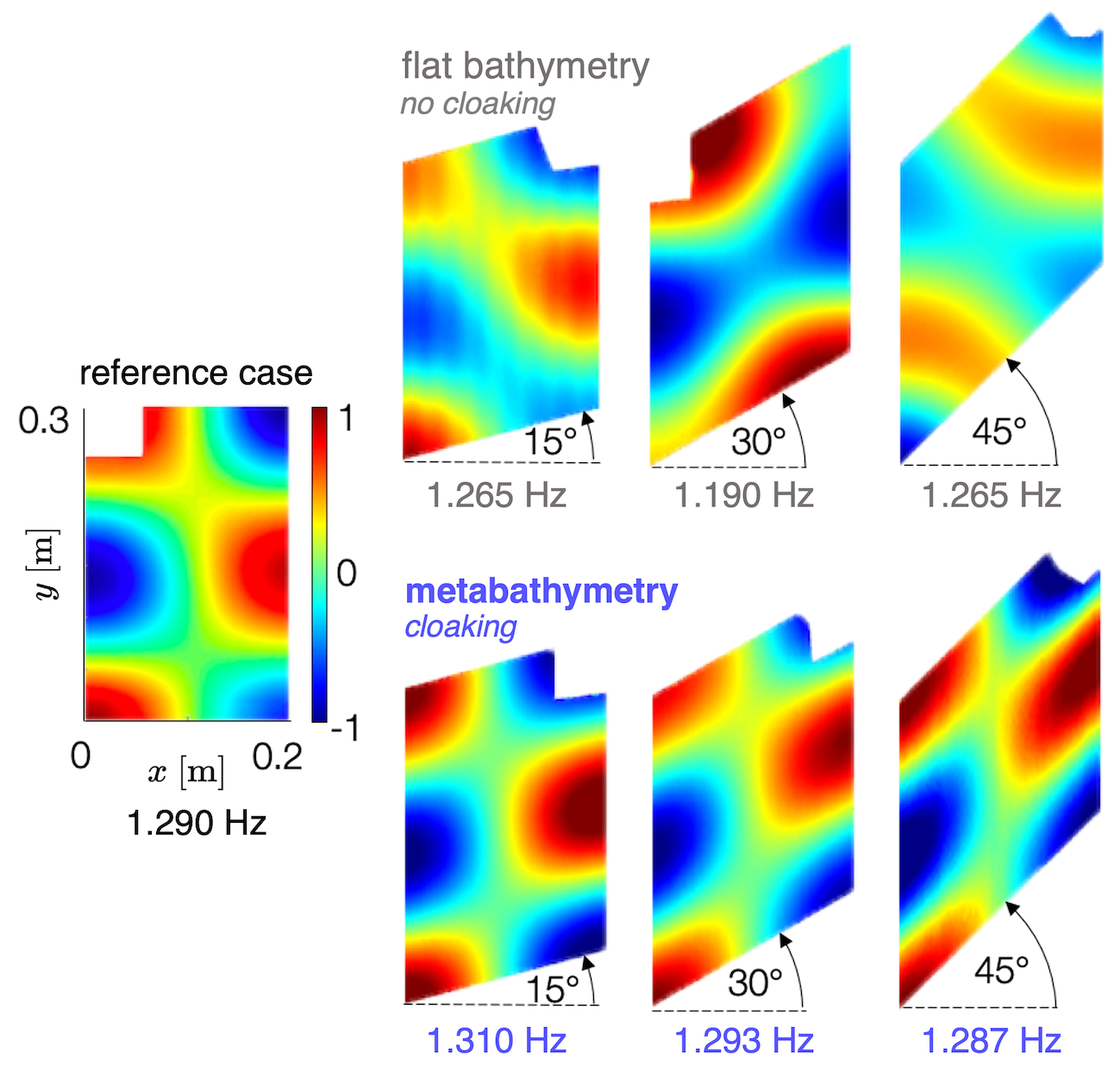} 
                \caption{\label{fig:resultsMode5} Experimental FTP measurements of the elevation of the free surface of the fifth eigenmode $\tilde\eta(x, y, \omega)$ for shear angles $\varphi=15^\circ$, $30^\circ$ and $45^\circ$ for the flat bathymetry and for the metabathymetry. The reference cavity for $\varphi=0^\circ$ is shown on the left panel.}
                \end{figure} 
                \begin{figure}[t]
                \includegraphics[width=.38\textwidth]{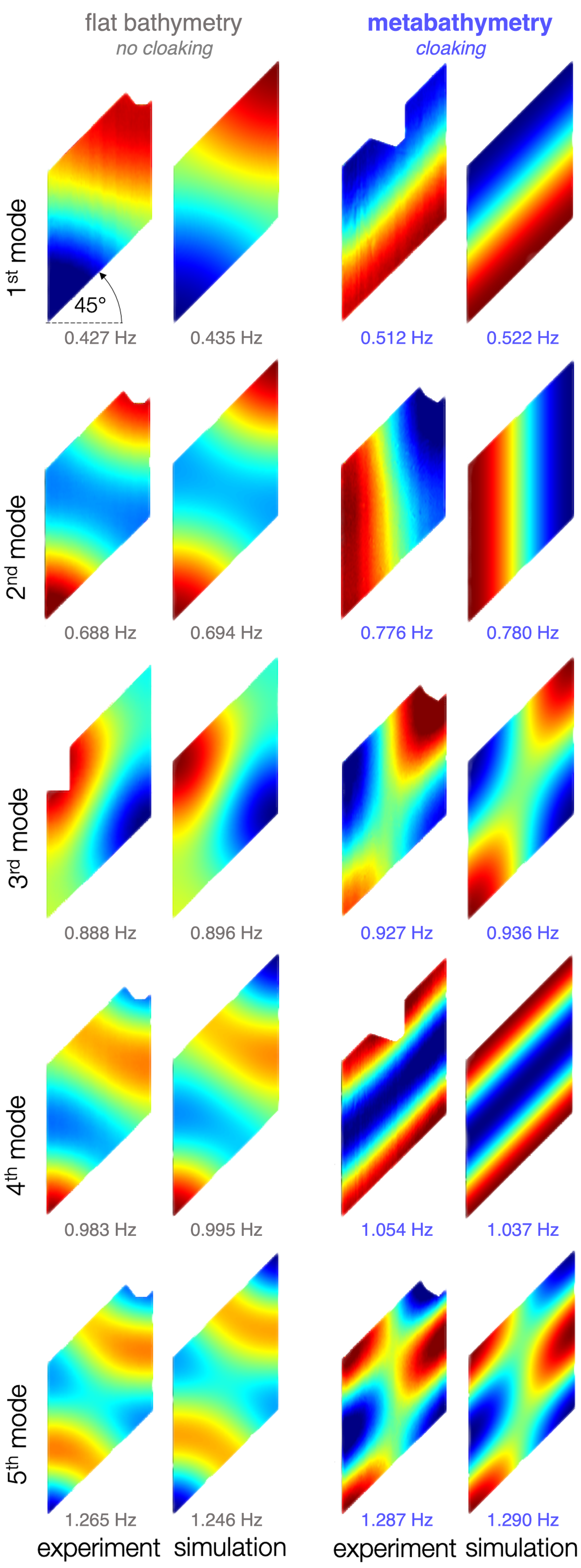} 
                \caption{\label{fig:results45} First five eigenmodes for $\varphi=45^\circ$. Wavefields $\tilde\eta(x, y, \omega)$ from FTP measurements and from numerical simulations, for the metabathymetry and for the flat bottom.}
                \end{figure}

Three metamaterial cavities with shear angles $\varphi_1=15^\circ$, $\varphi_2=30^\circ$, and $\varphi_3=45^\circ$ were designed and manufactured using a rapid prototyping technique (Fig. \ref{fig:photoExp}d-e). Three reference cavities with flat bathymetry and of the same shear angles $\varphi_1$, $\varphi_2$, and $\varphi_3$ were also built. The dimensions of the cavities $L_x$, $L_y$ were constant to preserve the same volume throughout the set of cavities and were set to $L_x=200 \, \mathrm{mm}$ and $L_y=300 \,\mathrm{mm}$. The reference water depth was chosen as $h_0=10 \, \mathrm{mm}$ as a trade-off between staying close to the shallow water regime and avoiding undesirable attenuation caused by friction in the bottom for small water depths. For each of the three systems, we calculate the parameters $h_{x_\ell}$, $h_{y_\ell}$, $\alpha$, and later $h^+$, $h^-$ based on the routine presented before, and we summarize them in Table \ref{tab:table1}. Considering the wavemaker constraints ($h^- \geq 3 \, \textrm{mm}$) and the rigidity limitations of the 3D printed structure ($\theta \ell  \geq 1 \, \textrm{mm}$) we chose $\theta = 0.2$ and $\ell = 5 \, \textrm{mm}$.

The partially submerged point source generates vertical motion with a 2 mm amplitude using a chirp signal spanning 0.3 Hz to 1.5 Hz for 40 seconds, enabling recovery of the first five cavity eigenmodes. The placement of the source is selected based on anticipated eigenmode shapes to ensure effective excitation. Occasional adjustments are necessary to prevent placement at eigenmode nodes, which would impede recovery. The placement of the wavemaker is sufficiently isolated from measurement regions to minimize near-field effects.

In order to quantify the wavefield, we use the Fourier Transform Profilometry (FTP) technique \cite{cobelli2009global} as well as confocal displacement sensors (2 lasers Keyence CL-P070). FTP is a technique that uses a fringe pattern projection on a measured surface. In our case, the water is painted with titanium dioxide ($\mathrm{TiO}_2$), so that its surface becomes diffusive and ready for a fringe projection whereas the change of physical properties of the painted water, including viscosity and surface tension, is insignificant \cite{Przadka2012}. A high-resolution video projector EPSON EH-TW9200W is used to project the fringe pattern and a high-speed camera Photron FASTCAM Mini WX100 records the deformation of the surface. Confocal displacement sensors are used separately to measure the amplitude of the wave in the maxima of eigenmodes with much higher accuracy than FTP and with pure, transparent water confirming the results of the FTP technique. Using these methods we obtain the space-time resolved measurements $\eta(x,y,t)$ that are later transformed into the frequency domain, resulting in the complex wavefield $\tilde{\eta}(x,y,\omega)$, to extract eigenfrequencies and eigenmodes. 
        
                \begin{table}[b]
                \caption{\label{tab:table1}Cavities dimensions and design parameters.}
                \begin{ruledtabular}
                \begin{tabular}{rrrr}
                \textrm{$\varphi$ [$^\circ$]}&
                \textrm{15}&
                \textrm{30}&
                \textrm{45}\\
                \colrule
                $\alpha$ [$^\circ$] & -41.18 & -36.95 & -31.72\\ 
                $h_{x_\ell}$ [mm] & 7.66& 5.66 & 3.82\\
                $h_{y_\ell}$ [mm] & 13.06 & 17.68& 26.18\\
                $h^+$ [mm] & 14.56 & 20.83 & 31.92\\
                $h^-$ [mm] & 7.07 & 5.07 & 3.23\\
                \end{tabular}
                \end{ruledtabular}
                \end{table}
                
                \begin{figure}[t]
                \includegraphics[width=.48\textwidth]{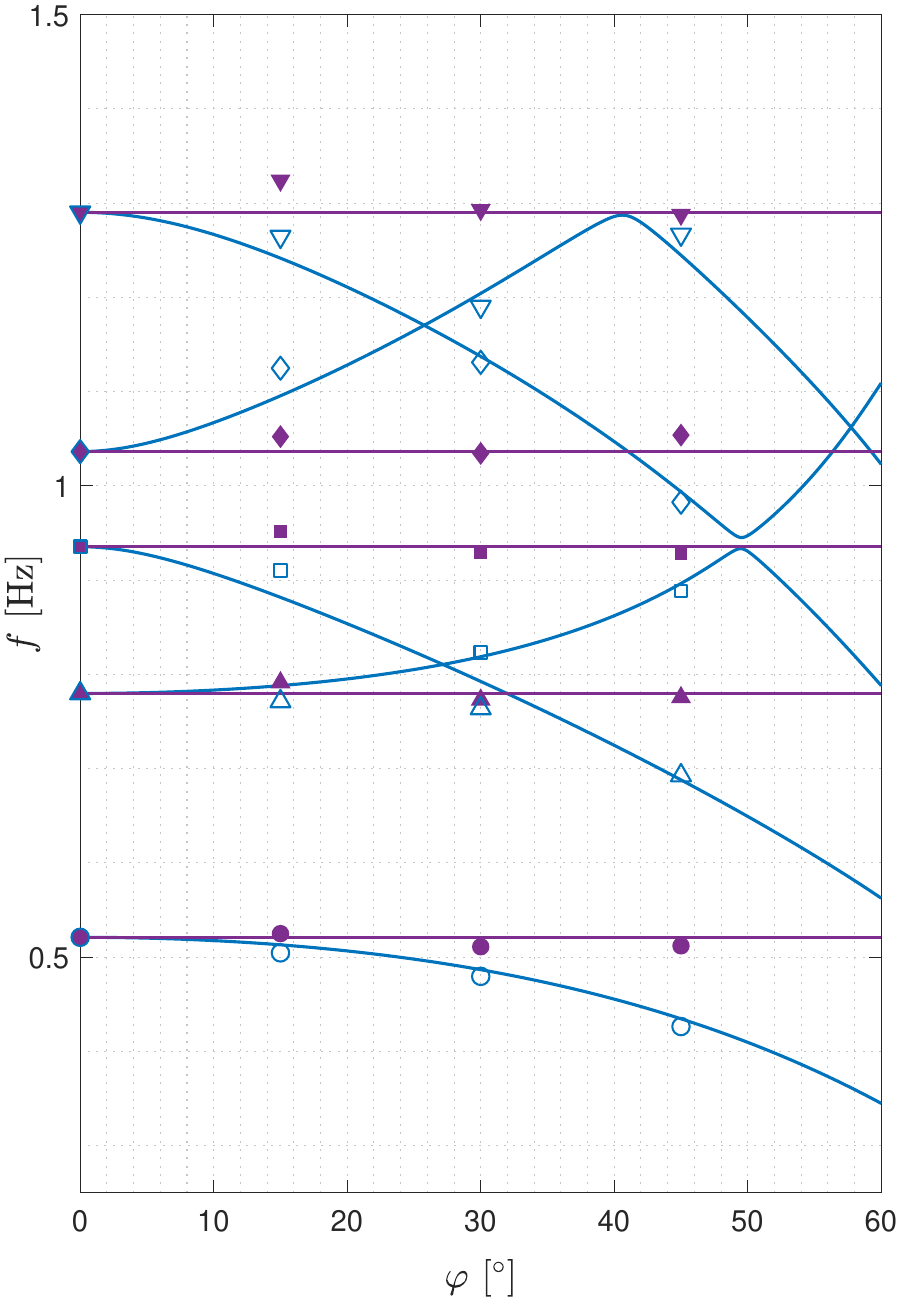} 
                \caption{\label{fig:resultsLaser} Experimental (symbols) and numerical (plain lines) values of the eigenfrequencies as a function of the shear angle $\varphi$. Results are shown for the anisotropic bathymetry (purple) and for the flat bathymetry (blue).}
                \end{figure} 

The selection of eigenfrequencies from the experiments is made by finding the local maxima of $|\tilde{\eta}(x,y,\omega)|$ averaged in space. Then, the eigenmodes are chosen as the real part of $\tilde{\eta}(x,y,\omega)$ at the given eigenfrequency and some of them are presented in Fig.~\ref{fig:resultsMode5} and Fig.~\ref{fig:results45}. The real part of the wavefield for the fifth eigenmode, i.e., the one whose frequency is the highest in the set of the measurements, is presented in Fig.~\ref{fig:resultsMode5} for rectangular reference cavity and cavities with the shear angles $\varphi$ of $0^\circ$, $15^\circ$, $30^\circ$ and $45^\circ$. In the reference cavities without the metabathymetry, the effect of the difference in geometry is clearly visible. The eigenmodes change their shape significantly with respect to the angle $\varphi$. The change in the position of the nodes and maxima is easily noticeable. The introduction of metabathymetry has an anticipated consequence. The eigenmode pattern remains the same throughout the set of measurements even for the highest angle $\varphi=45^\circ$. It is worth mentioning that in this case, i.e., the mode with the highest frequency and the highest shear angle, the pattern remains the same, even though the shallow water approximation here is questionable ($h^+=31.92 \, \mathrm{mm}$) and the friction of the metamaterial structure becomes more and more significant ($h^-=3.23 \, \mathrm{mm}$). This manifests a significant improvement and the benefit of using the homogenization of a fully three-dimensional linear water wave problem compared to previous works where 2D homogenization approach was used \cite{berraquero2013experimental}.

All the five eigenmodes recovered in this experiment for the highest angle $\varphi=45^\circ$ are reported in Fig.~\ref{fig:results45}. The higher modes were difficult to achieve for several reasons. First of all, as frequency increases, the eigenvalues become closer to each other, including degenerate cases, and are problematic if not impossible to distinguish experimentally. Moreover, when it comes to higher frequencies, dispersion and dissipation play a significant role in the water wave experiments, questioning the shallowness of the system and measurement techniques capacities, as the amplitude of the wave becomes extremely small. Note that increasing the amplitude and frequency of a water wave in our system would result in a nonlinear problem, which we do not study in this paper. 

Comparing the experimental results with the numerical prediction, obtained by solving an eigenvalue problem (for both regular and deformed cavities) using the Finite Element Method, we observe an excellent agreement of the eigenmodes (Fig.~\ref{fig:results45}). To quantitatively describe the difference between them, we introduce a pattern error defined as $ \epsilon_P\equiv \int_A |\eta_S(x,y,\omega) - \hat\eta(x,y,\omega) |^2 \, \mathrm{d}A / \int_A |\eta_S(x,y,\omega) |^2 \, \mathrm{d}A$, where $\eta_S$ is the normalized wavefield predicted numerically, $\hat\eta$ denotes the normalized wavefield measured in the experiments, and $A$ is the area of the cavity. Normalization is accomplished by rescaling the wavefield amplitude so that it satisfies the following condition $\int_A |\eta(x,y,\omega) |^2 \, \mathrm{d}A = 1$. The error increases with increasing angle $\varphi$ and frequency $\omega$. However, in all cases, it does not exceed $\epsilon_P=5.2\%$. The summary of all the experimental values of the eigenfrequencies compared to the numerical predictions is shown in Fig.~\ref{fig:resultsLaser}. The horizontal axis represents the angle of cavity deformation $\varphi$ and the vertical axis represents the frequency $f$. It can be seen that the use of metabathymetry allows to have a constant value of eigenfrequencies (purple dots), hence meeting our goal and preserving the resonance properties of the deformed cavities as predicted with the coordinate transformation theory. The efficiency of metabathymetry is quantified by the eigenfrequency error defined as $\epsilon_F=|f_S-f|/|f_S|$, where $f_S$ is the predicted eigenfrequency, and $f$ represents the eigenfrequency measured using confocal displacement sensors. This error is always smaller than the pattern error $\epsilon_P$, and its value varies from 1.4\% to 3.1\%. 

We have shown numerically and experimentally that a finely tuned metabathymetry allows to recover the resonance properties of a regular cavity in a deformed geometry. The efficiency of the anisotropic medium is confirmed using space-time resolved measurements of the full water wavefield. Very good agreement with numerical prediction is achieved. Such metabathymetries can be used to manipulate water wave resonances and they open perspectives in controlling sloshing inside irregular cavities.

\section{Acknowledgements}
       The authors acknowledge the support of the French Agence Nationale de la Recherche (ANR), under grant ANR-21-CE30-0046 (project CoProMM).
\section{Author declarations}
       The authors have no conflicts to disclose.
\section{Data availability}
       The data that support the findings of this study are available from the corresponding author upon reasonable request.
\bibliographystyle{aipnum4-1}
\bibliography{aipsamp}
\end{document}